\documentclass[12pt]{article}
\usepackage{simplemargins, amsmath, array, delarray, emlines2, epsfig,
newlfont}
\setallmargins{1in}  
\numberwithin{equation}{section}

\begin{document}

\thispagestyle{empty}
\begin{flushright}
hep-th/0312156
\end{flushright}
\vskip 2.5cm

\begin{center}\LARGE
{\bf AdS Solutions of 2D Type 0A}
\end{center}
\vskip 1.0cm
\begin{center}
{\large  David Mattoon Thompson\footnote{E-mail: {\tt thompson@physics.harvard.edu}}}

\vskip 0.5 cm
{\it Jefferson Physical Laboratory\\
Harvard University\\
Cambridge, MA  02138}
\end{center}

\vskip 0.8cm
\begin{center}
December 2003
\end{center}
\vskip 1.5cm

%%%%%%%%%%%%%%%%%%%%%%%%%% Abstract %%%%%%%%%%%%%%%%%%%%%%%%%%%%

\begin{abstract}
We present a two-parameter family of AdS solutions to the two-dimensional 
type 0A effective action.
\end{abstract}

\vfill
\newpage
\renewcommand{\baselinestretch}{2.2}

%%%%%%%%%%%%%%%%%%%%%%%% Introduction %%%%%%%%%%%%%%%%%%%%%%%%%%

\section{Introduction}

AdS backgrounds of string theory are often fruitful arenas for studying
holographic dualities and for constructing sigma models with R-R fluxes,
among other things.  AdS backgrounds of type 0A string theory should be no
exception.  In particular, the recent discovery of a matrix quantum mechanics dual to
two-dimensional type 0A string theory \cite{0307195, 0307083} suggests a promising direction
for understanding AdS$_2$/CFT$_1$ \cite{9711200}.  Also, two-dimensional
AdS presents one of the simplest backgrounds in which to study sigma models
in R-R flux.

As a first step in these pursuits, we present here a two-parameter family
of AdS$_2$ solutions to the two-dimensional type 0A effective action.  One
of the parameters is the tachyon vev $T$ (equivalently, the ratio of
dualized R-R field strengths, $q_+^2/q_-^2$).  The other parameter is the
string coupling $e^{2\Phi}$ (equivalently, the magnitude of the field
strength $q^2$).  In these solutions, string loops can be suppressed by
reducing the string coupling, but the high curvature of the spaces makes
higher order $\alpha'$ corrections important.

In section 2, we briefly review the 2D 0A string theory.  In section 3, we present our family of AdS$_2$
solutions.  In section 4, we discuss possible corrections to our solutions
from higher order terms in the effective action.

%%%%%%%%%%%%%%%%%%%%%%%% Review of 0A %%%%%%%%%%%%%%%%%%%%%%%%%%

\section{Two-Dimensional Type 0A}

The ten-dimensional type 0A string theory is given by the same worldsheet
action as the type IIA string, but with a GSO projection onto the closed
string sectors
\begin{equation}
\begin{array}{cccc}
\text{(NS+,NS+)}&\text{(NS$-$,NS$-$)}&\text{(R+,R$-$)}&\text{(R$-$,R+)}
\end{array}
\end{equation}
where $+$ and $-$ denote the eigenvalue of the worldsheet fermion number
operator $(-1)^F$.
In ten dimensions, each of these sectors contains a tower of states
corresponding to the possible transverse oscillations.  In two dimensions,
however, there is no room for transverse oscillations, so the situation is much
simpler.  We have the graviton $g_{\mu\nu}$ and dilaton $\Phi$ in the
(NS$+$,NS$+$) sector, the tachyon $T$ in the (NS$-$,NS$-$) sector, and two
gauge fields $C_\mu^{(\pm)}$ from the R-R sectors that give rise to two field
strengths, $F_{\mu\nu}^{(\pm)}$.  An allowed background for this theory is
the two-dimensional linear dilaton vacuum ($g_{\mu\nu} = \eta_{\mu\nu}$ and $\Phi =
\sqrt{\frac{2}{\alpha'}} \phi$) with an exponential tachyon wall ($T = \mu
e^{\sqrt{\frac{2}{\alpha'}}\phi}$) and zero field strengths
($F_{\mu\nu}^{(\pm)} = 0$).  The worldsheet action for this string theory is the
action of $\mathcal{N}=1$ super Liouville theory plus the action for a free
scalar superfield.

The action for $\mathcal{N} = 1$ super Liouville theory can be written in
superfield formalism\footnote{Unfortunately, it is standard in
the literature to use $\Phi$ for both the Liouville superfield and the
background dilaton.} as
\begin{equation}
S_{\text{SLT}} = \frac{1}{4\pi} \int d^2z d^2\theta \left( D\Phi
\overline{D} \Phi + 2i\mu e^{b\Phi} \right)\, ,
\end{equation}
where $\Phi$ is the scalar superfield
\begin{equation}
\Phi = \sqrt{\frac{2}{\alpha'}}\phi + i \theta \psi + i \overline{\theta}\overline{\psi} + i \theta
\overline{\theta} F\, ,
\end{equation}
and the covariant derivatives are given by
\begin{equation}
D = \partial_\theta + \theta \partial_z\, , \quad \overline{D} =
\partial_{\overline{\theta}} + \overline{\theta} \partial_{\overline{z}}\,
.
\end{equation}
In the case $b=1$, this yields a theory with central charge $\hat{c} = 9$.
When combined with the $\hat{c}=1$ theory of a free scalar superfield $X$ with
action
\begin{equation}
S_{\text{X}} = \frac{1}{4\pi} \int d^2z d^2\theta \left( DX \overline{D}
X \right)\, ,
\end{equation}
we get a critical SCFT with central charge $\hat{c} = 10$.  This is the
worldsheet action for two-dimensional type 0A in the linear dilaton vacuum.

The effective spacetime action was calculated in \cite{9811035} and was 
found to be
\begin{multline}\label{spacetimeaction}
\int d^2x \sqrt{-g} \left[ \frac{e^{-2\Phi}}{2\kappa^2} 
\left(\frac{8}{\alpha'} + R + 4(\nabla \Phi)^2 - f_1(T) (\nabla T)^2 + 
f_2(T) +  \ldots\right) \right. \\
\left. - \frac{\pi \alpha'}{2} f_3(T) \left(F^{(+)}\right)^2 - \frac{\pi 
\alpha'}{2} f_3(-T) \left(F^{(-)}\right)^2 + \ldots \right]\, .
\end{multline}
The first few terms in a Taylor expansion of the $f_i$ functions are
\begin{equation}
f_1(T) = \frac{1}{2} + \ldots\, , \qquad f_2(T) = \frac{1}{\alpha'} T^2 + 
\ldots\, , \qquad f_3(T) = 1 - 2T + 2T^2 + \ldots\, .
\end{equation}
There is evidence that the exact expression for $f_3(T)$ is $e^{-2T}$ 
\cite{0103244, 9901101}, and we will use this form for $f_3$ in our 
calculations.

%%%%%%%%%%%%%%%%%%%%%%%% AdS Solutions %%%%%%%%%%%%%%%%%%%%%%%%%

\section{AdS$_2$ Solutions}

\subsection{Equations of Motion}

To simplify the action (\ref{spacetimeaction}), we will dualize the 
R-R field strengths:
\begin{eqnarray*}
-\frac{2\pi\alpha^\prime}{4} f_3(\pm T) 
\left(F^{(\pm)}\right)^2 *1 &=& -\pi \alpha^\prime f_3(\pm T) F^{(\pm)} 
\wedge *F^{(\pm)}\\
&\longrightarrow& -\frac{1}{4\pi \alpha^\prime} f_3(\mp T) q_\pm^2 *1 + q_\pm 
F^{(\pm)} \\
&\longrightarrow& -\frac{1}{4\pi \alpha^\prime} q_\pm^2 f_3(\mp T) *1\, .
\end{eqnarray*}
In the second line, we have introduced an auxiliary field $q_\pm$.  The 
equation of motion for $q_\pm$ is
\begin{equation}
q_\pm = -2\pi\alpha' f_3(\pm T) *F^{(\pm)}\, ,
\end{equation}
which, when substituted in, gives the original action.  In the third line, 
we have integrated out $A^{(\pm)}$ which constrains $q_\pm$ to be a 
constant.  Therefore, in the third line, the fields $A^{(\pm)}$ and 
$q_\pm$ are no longer functionally integrated.  The full action can now be 
written as
\begin{multline}   
S = \int dx dt \sqrt{-g} \Bigg [
\frac{e^{-2\Phi}}{2\kappa^2} \left(
\frac{8}{\alpha'} + R + 4 (\nabla \Phi)^2  - f_1(T) (\nabla T)^2 + f_2(T)
+ \ldots \right)\\
- \frac{1}{4\pi \alpha'} f_3(-T) q_+^2 - \frac{1}{4\pi \alpha'} f_3(T)
q_-^2 + \ldots \Bigg ] \, .
\end{multline}

Varying with respect to the metric $g_{\mu\nu}$, dilaton $\Phi$, and
tachyon $T$ gives the equations of motion
\begin{multline}
\mathbf{(\delta g)} \qquad \frac{1}{2} g^{\mu\nu} \Bigg [ 
\frac{e^{-2\Phi}}{2\kappa^2} \left( \frac{8}{\alpha'} + 4 \nabla^2 \Phi - 4
(\nabla \Phi)^2 - f_1(T) (\nabla T)^2 + f_2(T)\right) \\
- \frac{1}{4\pi \alpha'} f_3(-T) q_+^2 
- \frac{1}{4\pi \alpha'} f_3(T) q_-^2 \Bigg ] \\
+\frac{e^{-2\Phi}}{2\kappa^2} \left(-2 \nabla^\mu \nabla^\nu \Phi + f_1(T)
\nabla^\mu T \nabla^\nu T \right) = 0\, ,
\end{multline}
\begin{equation}
\mathbf{(\delta \Phi)} \qquad 
\frac{8}{\alpha'} + R + 4 \nabla^2 \Phi - 4 (\nabla \Phi)^2 - f_1(T)
(\nabla T)^2 + f_2(T) = 0\, ,
\end{equation}
and
\begin{multline}
\mathbf{(\delta T)} \qquad \frac{e^{-2\Phi}}{2\kappa^2} \left[2f_1(T)
\nabla^2 T + f_1^\prime (T) (\nabla T)^2 - 4 f_1(T) (\nabla_\mu \Phi)
(\nabla^\mu T) + f_2^\prime(T) \right] \\
 - \frac{1}{4\pi \alpha'} f_3^\prime (-T) q_+^2 - \frac{1}{4\pi \alpha'} f_3^\prime (T) q_-^2 = 
0\, ,
\end{multline}
where primes denote differentiation with respect to $T$.
Setting $\Phi$ and $T$ constant, we find
\begin{equation}\label{gconstantequation}
\mathbf{(\delta g)} \qquad 
\frac{e^{-2\Phi}}{2\kappa^2} \left( 
\frac{8}{\alpha'} + f_2(T) \right) - \frac{1}{4\pi \alpha'} q_+^2 f_3(-T) - \frac{1}{4\pi 
\alpha'} q_-^2 f_3(T) = 0\, ,
\end{equation}
\begin{equation}\label{phiconstantequation}
\mathbf{(\delta\Phi)} \qquad \frac{8}{\alpha'} + R + f_2(T) = 0\, ,
\end{equation}
and
\begin{equation}\label{Tconstantequation}
\mathbf{(\delta T)} \qquad \frac{e^{-2\Phi}}{2\kappa^2} f_2^\prime (T) - 
\frac{1}{4\pi \alpha'} q_+^2 f_3^\prime(-T) - \frac{1}{4\pi \alpha'} q_-^2 f_3^\prime(T) = 0\, .
\end{equation}
With the $AdS_2$ metric
\begin{equation}
ds^2 = \frac{-4l^2}{\sin^2(u^+ - u^-)} du^+ du^-\, ,
\end{equation}
the Ricci scalar is
\begin{equation}
R = -2 / l^2\, .
\end{equation}

\subsection{Solutions $T=0$}

The solution with $q_- = q_+ \equiv q$ and $T=0$ satisfies the equations 
of motion with AdS radius given by 
\begin{equation}\label{radiusequation}
l^2 = \alpha'/4
\end{equation}
and dilaton given by
\begin{equation}\label{dilatonequation}
e^{-2\Phi} = \frac{\kappa^2}{8\pi} q^2\, .
\end{equation}
A notable feature of this solution is that the curvature radius is fixed at
a value of order the string length.  This implies that higher order
$\alpha'$ terms in the effective action will be important.  This will be
addressed in section 4.  Also, note that we are free to tune the string
coupling to zero by ramping up the strength of the R-R flux.

In this case, the ``tachyon'' is massive for all values of $q$.  
This can be seen as follows.  The $\delta T$ equation of motion, to first
order in $T$, gives us
\begin{equation}
\left\{ \nabla^2 + \nabla^2 \Phi - (\nabla \Phi)^2 + \frac{2}{\alpha'} - \frac{4\kappa^2}{\pi\alpha'} e^{2\Phi} q^2 \right\} \left(e^{-\Phi} T\right) =
0\, .
\end{equation}
The $\delta \Phi$ equation of motion, to zero order in $T$, tells us that
\begin{equation}
\nabla^2 \Phi - (\nabla \Phi)^2 + \frac{2}{\alpha'} = -\frac{R}{4}\, ,
\end{equation}
and, when substituted into the linearized $\delta T$ equation, gives us
\begin{equation}
\left\{ \nabla^2 - \frac{R}{4} - \frac{4\kappa^2}{\pi\alpha'} e^{2\Phi} q^2
\right\} \left(e^{-\Phi} T\right) = 0\, .
\end{equation}
Finally, substituting our background expressions for $\Phi$ and $R$, we get
\begin{equation}
\left(\nabla^2 - \frac{30}{\alpha'} \right) \left(e^{-\Phi} T \right) = 0\,
,
\end{equation}
so that the tachyon mass is $m_T^2 = \frac{30}{\alpha'} = \frac{15}{2l^2}$.
The authors of \cite{9811035} noted that, in ten dimensions, R-R flux could
stabilize the tachyon potential.  In our two-dimensional case, we see that
the R-R flux makes the otherwise-massless tachyon massive.

Solutions to the wave equation in AdS$_2$ are most readily attained in
Poincare coordinates, in which
\begin{equation}
ds^2 = l^2 \frac{-dt^2 + dy^2}{y^2}\, .
\end{equation}
In these coordinates, the wave equation takes the form
\begin{equation}
\left(\frac{\partial^2}{\partial y^2} - \frac{\partial^2}{\partial t^2} -
\frac{l^2 m_T^2}{y^2} \right) T(t,y) = 0\, .
\end{equation}
Using separation of variables, we can write the general time-dependent,
positive-frequency solution as $e^{-i\omega t} \chi(y)$.  The normalizable solution is readily obtained in
terms of a Bessel function as
\begin{equation}
T_w(t,y) = e^{-i\omega t} \sqrt{\frac{y}{2}} J_{h_\pm-1/2} (\omega y)\, ,
\end{equation}
where $h_\pm = \frac{1}{2} \pm \frac{1}{2} \sqrt{1+4l^2m_T^2}$.  The static
solutions are obtained by noting that the wave equation
\begin{equation}
y^2 \frac{\partial^2}{\partial y^2} T = l^2 m_T^2 T
\end{equation}
implies that $T \sim y^n$ where $n(n-1) = l^2 m_T^2$.  Therefore, the
general static solution is
\begin{equation}
T = a y^{h_+} + b y^{h_-}\, .
\end{equation}
Note that, although these static solutions are non-normalizable, they may
make an appearance as approximate solutions in regions of spacetime that
are AdS-like.

Solutions to the wave equation in global coordinates are a little more
difficult to come by, but they have been worked out in \cite{9812047}.  In
the global coordinates
\begin{equation}
ds^2 = l^2 \frac{-d\tau^2 + d\sigma^2}{cos^2\sigma}\, ,
\end{equation}
the normalized positive-frequency solutions are
\begin{equation}
T_n(\tau, \sigma) = \Gamma(h) 2^{h-1} \sqrt{\frac{n!}{\pi \Gamma(n+2h)}}
e^{-i(n+h)\tau} (\cos \sigma)^h C^h_n (\sin \sigma)\, ,
\end{equation}
where $n=0,1,2,\ldots$, $C_n^h$ is the Gegenbauer polynomial, and $h$ is
once again related to $m_T^2$ by $h(h-1) = l^2 m_T^2$.  Note
that, unlike in Poincare coordinates, the spectrum in global coordinates is
discrete.

\subsection{Solutions with $T \neq 0$}

The solution given in the previous section can be deformed by moving the
constant value of $T$ away from zero.
The solution is given by
\begin{equation}
l^2 = \frac{\alpha'/4}{1+\frac{\alpha'}{8} f_2(T)}\, ,
\end{equation}
\begin{equation}
e^{-2\Phi} = \frac{\frac{\kappa^2}{16\pi} \left(q_+^2 
f_3(-T) + q_-^2 f_3(T) \right)}{1+\frac{\alpha'}{8} f_2(T)}\, ,
\end{equation}
and
\begin{equation}
\frac{q_-^2}{q_+^2} = \frac{f_3(-T)}{f_3(T)} \frac{8/\alpha' + f_2(T) - 
f_2^\prime (T)/2}{8/\alpha' + f_2(T) + f_2^\prime(T) / 2}\, .
\end{equation}
Again, it is clear that, for all solutions in this family, we can send the string 
coupling to zero while holding fixed both the tachyon vev $T$ and the AdS 
radius $l$.  This is accomplished by sending $q_-^2$ and $q_+^2$ to 
infinity while holding the ratio $q_-^2 / q_+^2$ fixed.

It is not evident from these equations whether or not there exists an
AdS$_2$ solution with one of the $q$'s, say $q_-$, set to zero.  Setting
$q_-=0$ would require a $T$ of order 1, but to understand such large values
of $T$ would require a more complete knowledge of $f_2$.  Specifically,
$q_-=0$ would require that
\begin{equation}
\frac{8}{\alpha'} + f_2(T) - \frac{1}{2} f_2^\prime(T) = 0\, ,
\end{equation}
and it is not known whether this equation has solutions.

%%%%%%%%%%%%%%%%%%%%%%%%%% Discussion %%%%%%%%%%%%%%%%%%%%%%%%%%

\section{Discussion}

It should be noted that the AdS spaces presented here are solutions to the first few terms in the
effective action.  Since the AdS radius is of order the string length,
we expect higher order terms in $\alpha'$ to change some of the
quantitative features of the solutions, such as the exact value of the AdS
radius or the true mass of the tachyon.  However, as we shall discuss here,
the qualitative features of the AdS solutions are rather generic and are
not expected to be changed by the higher order $\alpha'$ terms.

We can ask what other terms might make contributions to the
equations of motion, and, therefore, might change features of the AdS
solution.  For simplicity, let us concentrate on the $T=0$ solution found
in section 3.3.  Since we seek a solution with $\Phi$ constant and $T$
zero, any terms linear or higher order in $\nabla \Phi$ or $T$ will make no
contribution to the $\delta g$ or $\delta \Phi$ equation of motion.  Therefore, the only
higher order NS-NS terms that will contribute are contractions of the
Riemann tensor.  
Since we are in
two dimensions, the Riemann tensor is related to the Ricci scalar as
\begin{equation}
R_{\mu\nu\alpha\beta} = \frac{1}{2} (g_{\mu\alpha}g_{\nu\beta} -
g_{\mu\beta} g_{\nu\alpha}) R\, .
\end{equation}
As a result, any scalar combination of $n$ Riemann tensors can be
rewritten as $b R^n$ for some constant $b$.  Therefore, without loss of
generality, we may consider replacing $R$ in the effective action with
$R+\sum\limits_{n=2}^\infty {\alpha^\prime}^{n-1} b_n R^n$.  In our $T=0$ solution, these terms will change 
equation (\ref{radiusequation}) to
\begin{equation}
\frac{8}{\alpha'} - \frac{2}{l^2} + \sum\limits_{n=2}^\infty
{\alpha^\prime}^{n-1} b_n \left(\frac{-2}{l^2}\right)^n = 0\, .
\end{equation}
So long as there are real roots to this equation, then there will exist an
AdS solution with the corresponding radius.
The equation
(\ref{dilatonequation}) for the string coupling is modified to
\begin{equation}
e^{-2\Phi} = \frac{\frac{\kappa^2}{8\pi} q^2}{1 - \frac{1}{8}
\sum\limits_{n=2}^\infty (n-1) {\alpha'}^n b_n \left(\frac{-2}{l^2}\right)^n}\, .
\end{equation}
Once again, it is clear that the string coupling may be taken to zero by
making $q_+$ large.

The $\delta T$ equation of motion is, of course, unaffected by these $R^n$
terms.  On the other hand, the $\delta T$ equation will be affected by
terms in the action that are linear in $T$.  However, there are no NS-NS
terms involving odd powers of $T$ \cite{9811035}.  Terms with odd powers of
$T$ are possible when they also involve R-R fields, for example, the
$e^{\pm 2T} q_\pm^2$ terms.  However, the proposed symmetry \cite{0307195}
of the theory under $T \rightarrow -T$ and $F^{(+)} \leftrightarrow
F^{(-)}$ would imply that setting $q_+ = q_-$ makes such terms zero.

We may also ask how the tachyon mass would be altered by the terms higher
order in $\alpha'$.  For example, NS-NS terms of the form $R^n T^2$ would
change the tachyon mass.  In fact, we fully expect the value of the tachyon
mass to be corrected by higher order terms.

It is clear that the form of the AdS solutions are rather generic and are
not likely to be dramatically changed by terms higher order in $\alpha'$.
This fact motivates a search for the corresponding worldsheet sigma model
describing type 0A strings propagating in these AdS$_2$ spaces.  Because of
the existence of nonzero R-R fluxes, the correct sigma model will most
likely not be found using the NSR formalism.  Fortunately, several other
worldsheet formalisms have been developed that have allowed for
quantization of the string in R-R backgrounds.  For example, the hybrid
formalism has been used to study superstring quantization in AdS$_3$ $\times$ S$^3$
\cite{9902098}, AdS$_2$ $\times$ S$^2$ backgrounds \cite{9907200}, and
curved 2D backgrounds \cite{0107140}.

%%%%%%%%%%%%%%%%%%%%%% Acknowledgements %%%%%%%%%%%%%%%%%%%%%%%%

\section{Acknowledgements}

This work was supported in part by DOE grant DE-FG02-91ER40654.  I am
grateful to A. Simons and A. Strominger for helpful conversations.  I would
also like to thank my friends at Eastern Electronics for their last-minute
technical support.

%%%%%%%%%%%%%%%%%%%%%%%% Bibliography %%%%%%%%%%%%%%%%%%%%%%%%%%  

\end{document}